\documentclass[12pt,epsfig,epsf]{article}
\usepackage{cite}
\usepackage{graphicx} 
\usepackage{amsfonts} 
\usepackage{amssymb} 

\def\beqa{\begin{eqnarray}}
\def\eeqa{\end{eqnarray}}
\def\lsim{\mathrel{\raise.3ex\hbox{$<$\kern-.75em\lower1ex\hbox{$\sim$}}} }
\def\gsim{\mathrel{\raise.3ex\hbox{$>$\kern-.75em\lower1ex\hbox{$\sim$}}} }

\begin{document}

\hfill{} 
\begin{tabular}{l} 
MPP-2003-42\\PM 03-21\\
\vspace*{0.4cm}
\end{tabular} 

\newcommand{\nn}{\noindent}
\newcommand{\nnu}{\nonumber}
\renewcommand{\thefootnote}{\fnsymbol{footnote}}
\begin{center}
{\Large{ {\bf Higgs decays in the Two Higgs Doublet Model: \\
Large quantum effects in the decoupling regime}}}

\vspace{.8cm}

{\large A. Arhrib}$^{\mbox{1,2}}$,
{\large M. Capdequi Peyran\`ere}$^{\mbox{3}}$,
{\large W. Hollik}$^{\mbox{1}}$, {\large S. Pe\~naranda}$^{\mbox{1}}$,
\vspace{.6cm}

1: Max-Planck-Institut f\"ur Physik (Werner-Heisenberg-Institut) \\
F\"ohringer Ring 6, D--80805 Munich, Germany\\ 

\vspace{.6cm}
2: D\'epartement de Math\'ematiques, Facult\'e des Sciences et Techniques\\
B.P 416 Tanger, Morocco.\\
and\\
LPHEA, D\'epartement de Physique, Facult\'e des Sciences-Semlalia,\\
B.P. 2390 Marrakesh, Morocco.

\vspace{.6cm}
3: Laboratoire de Physique Math\'ematique et Th\'eorique, CNRS-UMR 5825\\
Universit\'e Montpellier II, F--34095 Montpellier Cedex 5, France

\end{center}

\setcounter{footnote}{4}
\vspace{1.2cm}

\begin{abstract}
We study the Higgs-boson decays 
$h^0\to b\bar{b}$, $h^0\to \gamma\gamma$ and $h^0\to \gamma Z$ 
within the framework of the Two Higgs Doublet Model (THDM) in the 
context of the decoupling regime, together with tree level unitarity 
constraints. We show that when the light CP-even Higgs boson of 
the THDM mimics the Standard-Model Higgs boson, not only the 
one-loop effects to $h^0\to \{ \gamma\gamma ,\gamma Z \}$
but also the one-loop contribution to $h^0\to b\bar{b}$ 
can be used to distinguish between THDM and SM. The size 
of the quantum effects in $h^0\to b\bar{b}$ are of the same 
order as in $h^0\to \{ \gamma\gamma ,\gamma Z \}$ and can reach 
25\% in both cases.
\end{abstract}

\vfill

\newpage

\renewcommand{\thefootnote}{\arabic{footnote} }
\setcounter{footnote}{0}
\section{Introduction}
The discovery of a Higgs boson is one of 
the major goals of present and future searches in particle physics. 
Global electroweak fits within the Standard Model (SM) 
yield an upper bound on the Higgs-boson mass of
$M_H < 211$ GeV at $95\%$ confidence level (CL)~\cite{grun}. 
Together with the direct-search limit from the
LEP experiments~\cite{lep} of $M_H > 114.4$~GeV,
the Higgs boson seems ``just around the corner".

The still hypothetical Higgs sector of the SM can be 
enlarged and some simple extensions such as the 
Two Higgs Doublet Model (THDM) versions~\cite{gun} are 
intensively studied. 
Such extensions add new phenomena, like the charged 
Higgs-boson sector, and satisfies the relevant constraint 
$\rho \approx 1$ up to finite radiative corrections.
Actually, there are two versions of the THDM, type I and type II, 
differing in some Higgs couplings to fermions, but in both 
types and after electroweak symmetry breaking, 
the Higgs spectrum is the same: two charged Higgs particles
$H^\pm$, two CP-even $H^0$, $h^0$ and one CP--odd $A^0$. 
Aside the charged Higgs sector, the neutral sector of the THDM 
is noticeably different from the SM single-Higgs sector. 
Whereas the discovery 
of a charged scalar Higgs boson would attest definitively that the 
Standard Model is overcome, more refined investigations would be
necessary in case of a neutral Higgs particle, in the SM
and beyond~\cite{beySM}. 

In order to establish 
the Higgs mechanism for the electroweak symmetry breaking, 
we need to measure the Higgs 
couplings to fermions and to gauge bosons as well as the self-interaction 
of Higgs bosons. Such measurements, if precise enough,  
can be helpful in discriminating between the models
through their sensitivity to quantum-correction effects,
in particular in specific cases like the decoupling limit.
For a  Linear Collider,
it has been shown~\cite{marco} that the Higgs-boson couplings 
to third-generation fermions and gauge bosons can be measured
with high precision of the order 1--3\%.

The {\it decoupling theorem}~\cite{ac} states, in brief, that,
if we have a theory with light and heavy particles,
under certain conditions, the effects of heavy particles only appear in
the low-energy theory through renormalization of couplings and
masses of the effective theory
or through corrections proportional to a negative power of the
heavy masses. The {\it decoupling theorem}
is not universal and suffers from some known exceptions~\cite{col}.
To be sure that the theorem holds, the low-energy theory has to be
renormalizable, and it should have neither spontaneous symmetry breaking nor
chiral fermions, which does not apply to the  MSSM and the THDM.
A formal and general proof of decoupling of the non-standard MSSM particles 
from the low-energy electroweak gauge-boson physics has been given
in~\cite{TesisS}.
Conversely, concerning Higgs physics, it is known that the SUSY one-loop
corrections do not decouple, in general, in the limit of a heavy
supersymmetric spectrum~\cite{dec-prop,Sola,herrero,ITP1,
madrid,siannah,okada}. 

Furthermore, it is not yet rigorously proven 
that this decoupling theorem applies in the case of the 
THDM \cite{howie}. But one can be less ambitious and consider a
weaker version for the {\it decoupling limit} \cite{hojon}, where all 
the scalar masses with one exception formally become infinite.
For the case of the THDM, this limit designs
the CP-even $h^0$ as the light scalar particle
while the other Higgs particles, the CP-even $H^0$, 
the CP-odd $A^0$, and the charged Higgs boson $H^{\pm}$ 
are extremely heavy and mass-degenerate. 
In using  pure algebraic arguments at the tree level, 
and more sophisticated ones at the loop level, one can derive 
the main consequences: in the {\it decoupling limit}, 
$\cos(\beta-\alpha)\to 0$, the CP-even $h^0$ of the THDM and 
the SM Higgs $H$ have quite similar tree level 
couplings to gauge bosons and fermions as well \cite{howie}.
Here $\alpha$ is the mixing angle in the CP-even sector
and $\tan \beta = v_1/v_2$, the ratio of the two vacuum 
expectation values\cite{gun}.
Since the scalar sector of the MSSM is a particular 
case of THDM, the various decoupling scenarios also 
take place in the MSSM and give similar consequences as in the THDM.

Obviously, the {\it decoupling limit} does not rigorously 
apply to a more realistic 
world where the particles masses are in a finite range. 
Actually, one may consider a less rigorous scenario, labeled as the  
{\it decoupling regime}~\cite{howie,hojon,Haberdec}, 
where we only require that the heavy 
Higgs particles  have masses much larger than the $Z$ boson mass. 
However, such a scenario with large masses has to 
agree with the 
unitarity of the $S$-matrix. The unitarity constraint puts in turn 
bounds on the amplitude of partial waves~\cite{unit}, which give finally 
constraints on the values of the coupling constants. The problem 
in models with symmetry breaking is that couplings and masses are not 
independent; hence, unitarity of the $S$-matrix implies upper bounds 
on masses in the SM as well as in the THDM.

Since perturbative expansion is used, it is impossible to 
find the exact bounds; instead,
one can derive tree-level unitarity 
bounds or loop-improved unitarity bounds.
In this study, we will use unitarity bounds coming from a tree-level 
analysis~\cite{abdesunit}. This tree level analysis is derived with 
the help of the equivalence theorem \cite{eqthe}, which itself is a 
high-energy approximation where it is assumed that the 
energy scale is much larger than the $Z^0$ and $W^\pm$ 
gauge-boson masses. We will consider here
this ``high-energy'' hypothesis that both the 
equivalence theorem and the decoupling regime 
are well settled, but in such a way that the  unitarity 
constraint is also fulfilled. Our purpose is to investigate 
the quantum effects in the decays of the light CP-even Higgs 
boson $h^0$, especially looking for sizeable differences with 
respect to the SM in the decoupling regime.

Several studies have been carried out 
looking for non-decoupling effects in Higgs-boson decays and 
Higgs self-interactions. Large loop effects in $h^0\to \gamma \gamma$
and $h^0\to \gamma Z$ have been pointed out for both the MSSM~\cite{ITP1}
and the THDM~\cite{maria}. The one-loop 
SUSY-QCD effects in $h^0\to b\bar{b}$ were addressed in~\cite{madrid}
and their decoupling properties are well discussed and understood.
The loop contribution
to the triple $h^0$ self-coupling has been investigated 
in the MSSM~\cite{siannah}
as well as in the THDM \cite{okada}, revealing non-decoupling effects
which, however, in the MSSM case disappear when the self-coupling 
is expressed in terms of the $h^0$ mass~\cite{siannah}.

The present study assumes the following chronology and scenario: 
the SM model is not ruled out by any experiment, no SUSY evidence, 
no charged Higgs and no CP-odd Higgs signal, a light CP-even 
Higgs is detected and its properties are rather close to the SM Higgs. 
A very natural question emerges: what kind of Higgs we got? 
It seems difficult to disentangle the Higgs particle of 
the basic SM from other Higgs bosons involved in extended models 
like the THDM or MSSM.  
We will focus on the perhaps most difficult scenario, where all 
the Higgs particles of the THDM, except the lightest CP-even Higgs, 
are heavy to escape detection at the first stage
of next generation colliders. 

In this paper we investigate three decay modes of the CP-even neutral Higgs 
boson $h^0$ (THDM) or $H$ (SM): the decay into a pair of photons, 
the decay into a photon in association with a $Z$ boson, 
and the decay into a $b\bar{b}$ quark pair.
The decays into the gauge-boson pairs are loop-mediated 
processes since the photon does not couple to neutral particles.
In contrast, the decay 
into a fermion pair already exists at the tree level because of
the Higgs--$b$ Yukawa interaction.
For the specific scenario where only a single light Higgs particle exists
with tree-level  coupling constants as in the SM,
the coupling structure of the virtual
non-standard heavy particles in the quantum contributions to the   
bosonic decay rates yield sizeable differences to the SM decay rates.
Simultaneously, these quantum corrections also influence the fermionic 
decay rates and thus make them differ significantly from the SM result
as well.  

The paper is organized as follows. 
In section 2 we address the Higgs-boson decay channels under study
and outline the calculation of the
one-loop contribution to the $h^0\to b\bar{b}$ partial width, giving
details about the renormalization scheme used.
Section 3 is devoted to the presentation and discussion of
our numerical results.

\section{Decays of the ${\bf \boldmath{h^0}}$ boson in the THDM}

In this section, we first discuss the one-loop contributions 
to $h^0\to \gamma\gamma$ and $h^0\to \gamma Z$, which have been known
already for a while~\cite{gun}.
Then, we present in more details the one-loop contribution to 
$h^0\to b\bar{b}$ as well as details of the renormalization scheme used.

We first start with the decay channels
$h^0\to \gamma\gamma$ and $h^0\to \gamma Z$, which are loop-induced,
the only pure THDM contribution comes from 
charged Higgs loops, which may lead to non-decoupling effects. 
In the decoupling regime, with $\alpha \rightarrow \beta-\pi/2$, 
the charged Higgs contribution enters at the one-loop level
through the following coupling,
\beqa
g[h^0H^+H^-] & = & 
-\frac{g}{2M_W}\{ M_{h^0}^2 +2(M_{H^\pm}^2 -\lambda_5 v^2)  \}
\label{h0hphm}
\eeqa
where $v^2=v_1^2 +v_2^2=(2\sqrt{2} G_F)^{-1}$ .
To derive this coupling we have used the CP conserving scalar potential
of ref.~\cite{gun}, where the parameter $\lambda_5$ breaks softly
the discrete symmetry $\Phi_1\to -\Phi_1$.
The decay rates of  $h^0\to \gamma\gamma$ and $h^0\to \gamma Z$ 
are taken from ref.~\cite{gun}.
From the form of the coupling in eq.~(\ref{h0hphm}),
one can see that the quadratic term $M_{H^\pm}^2$ can be compensated 
by $\lambda_5 v^2$. With charged Higgs-boson masses much larger than 
the electroweak scale ($M_{H^\pm}>v$), such cancellations take place only
for large values of $\lambda_5$ \cite{hojon}. 
For fixed $M_{h^0}$ and $M_{H^\pm}$, the charged Higgs contribution,
entering through eq.~(\ref{h0hphm}), vanishes
for the critical choice of $\lambda_5 = \lambda_5^0$, where 
\beqa
\lambda_5^0 = \frac{1}{2 v^2} \{ M_{h^0}^2 + 2 M_{H^\pm}^2 \}\label{zero}.
\eeqa

For the Higgs decay $h^0\to b\bar{b}$, 
we evaluate the partial width $\Gamma({h^0\to b\bar{b}})$
at the one-loop level,
in both the SM and the THDM, using the SM width as a reference.
The model-independent contributions, QCD and QED corrections, are not included.
For $\Gamma({H \to b\bar{b}})$ in the SM,
we have performed the calculation in the on-shell scheme~\cite{HBS}, 
in analogy to the work presented in~\cite{HDK}.
Practical  computations were  done with the help of the packages 
{\it FeynArts, FormCalc}~\cite{FA}, and with
LoopTools and FF for numerical evaluations~\cite{FF}.
The THDM is nowadays available in the package {\it FeynArts}.

At one-loop order the amplitude can be written as follows,
\beqa
\label{amplitude}
{\cal M}_1 = -\frac{ig m_b }{2 M_W c_\beta} \,
\sqrt{Z_{h^0}} \left[ s_{\alpha} \,  
(1 + \Delta {\cal M}_1 )  +  s_{\alpha}\, \Delta {\cal M}_{12} \right]
\eeqa 
with $s_\alpha = \sin\alpha\,, c_\beta = \cos \beta$, and
\beqa
\label{oneloopterms}
\Delta {\cal M}_1 &= &V_1^{h^0 b\bar{b}} + \delta (h^0 b\bar{b})\,\,\, , 
 \,\,\, \Delta {\cal M}_{12} = \frac{\Sigma_{hH}(M_{h^0}^2)}
                              {M_{h^0}^2 - M_{H^0}^2} 
                         -\delta\alpha \, , \nnu \\
Z_{h^0} &=& \left[ 1+\widehat{\Sigma}_{h^0}^{\prime}(M_{h^0}^2) 
            \right] ^{-1} \, .
\eeqa
These expressions contain the vertex corrections $V_1^{h^0 b\bar{b}}$
with the corresponding vertex counterterm $\delta (h^0 b\bar{b})$,
the non-diagonal $h^0$--$H^0$ self-energy $\Sigma_{hH}$ with
the counterterm $\delta\alpha$ for the mixing angle $\alpha$,
and the wave-function renormalization with the
$h^0$ field-renormalization constant $Z_{h^0}$ 
derived from the renormalized self-energy of the $h^0$.

In the general THDM, 
the mixing angle $\alpha$ is an independent parameter and can hence 
be renormalized in a way independent of all the other renormalization
conditions. A simple and natural condition is to require that $\delta\alpha$ 
absorbs the $h^0$--$H^0$ transition in the non-diagonal part 
$\Delta {\cal M}_{12}$ of the decay amplitude~(\ref{amplitude}). 
The angle $\alpha$ is hence the CP-even 
Higgs-boson mixing angle also at the one-loop level, and the
decay amplitude ${\cal M}_1$ simplifies to the $\Delta{\cal M}_1$ term only. 
The Higgs-boson decay width is then given by the expression
\beqa
\Gamma_1(h^0 \to b\bar{b}) = 
\frac{N_C G_F m_b^2}{4 \sqrt{2} \pi} 
\frac{s_\alpha^2}{c_\beta^2}\,
M_{h^0} \, Z_{h^0} \,[1 + 2 \Re (\Delta {\cal M}_1) ] \,.
\label{width}
\eeqa

\begin{figure}[t!] 
      \vspace*{-6.cm}         
       \centerline{
      \includegraphics*[scale=0.85]{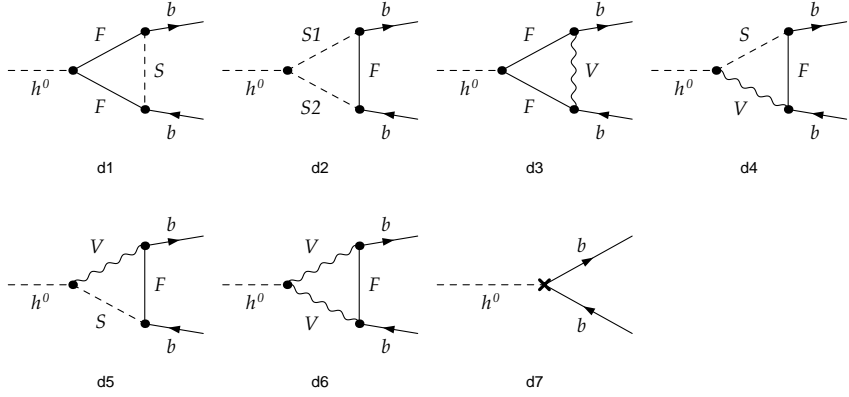} }
\vspace*{-13.5cm}
\caption{Generic one-loop THDM Feynman diagrams contributing to 
$\Gamma({h^0\to b\bar{b}})$.} 
\label{fig:diagrams}
\end{figure}

The central part of the computation is thus the determination
of $\Delta {\cal M}_1$.
The generic THDM contributions to $\Delta {\cal M}_1$
are depicted in Fig.~\ref{fig:diagrams}, 
involving vertex-correction and counterterm diagrams.
A quick inspection shows 
that there are  pure THDM contributions not present
in the SM case: diagram 
$d_1$ with $S=H^0,A^0,H^\pm$, diagram $d_2$ with $(S_1,S_2)=(H^0,H^0)$,
$(A^0,A^0)$, $(H^\pm,H^\pm)$ and $(H^\pm,G^\pm)$, and lastly diagrams $d_{4,5}$
with $S=A^0,H^\pm$. In the decoupling limit, 
diagrams $d_{4,5}$  and diagram $d_2$ with $(S_1,S_2)$=$(H^\pm,G^\pm)$  
vanish. As a consequence, large effects in 
$h^0\to b\bar{b}$ may arise from diagrams $d_1$ and $d_{2}$.
Thereby, $d_2$ formally has non-decoupling behaviour, yielding a non-zero
value in the mathematical limit of heavy non-standard particles.

We refrain here from giving analytical
expressions for the vertex corrections; the MSSM formulae
given in \cite{fermionic}
can be adapted to the THDM case replacing the MSSM couplings 
by the THDM ones~\cite{hojon,couplings}.

We will use the on-shell scheme based on~\cite{dabelstein}
for determination of the counterterms, with the exception that 
the field renormalization constants for the two Higgs doublets are 
determined in the  ${\overline{\rm MS}}$ scheme, yielding
$Z_{\Phi_i} = 1 + \delta Z_{\Phi_i}^{\overline{\rm MS}}$, 
\beqa
\label{Zfactors}
\delta Z_{\Phi_1}^{\overline{\rm MS}} & = &
\frac{-g^2 \Delta }{32 \pi^2 m_W^2 } \{ \frac{1}{c_\beta^2}
(m_e^2 + m_\mu^2 + m_\tau^2 + (m_b^2 + m_d^2 + m_s^2) N_C)\} 
+\frac{\Delta}{2(4\pi)^2} (3 g^2 + g^{\prime 2})\, , \nnu \\
\delta Z_{\Phi_2}^{\overline{\rm MS}} & = &
\frac{-g^2 \Delta }{32\pi^2 m_W^2} 
\{ \frac{N_C}{s_\beta^2} (m_c^2 + m_t^2 + m_u^2) \} 
+\frac{\Delta}{2(4\pi)^2} (3 g^2 + g^{\prime 2}) \, ,
\eeqa
with $\Delta =2/(4-D)-\gamma+{\rm log} 4\pi $ from dimensional
regularization and the color factor $N_C$. Accordingly,
the $h^0$ field-renormalization constant is a linear combination
of~(\ref{Zfactors}), which determines the derivative of the renormalized 
self-energy in~(\ref{oneloopterms}) to be
\beqa
\widehat{\Sigma}_{h^0}^{\prime}(M_{h^0}^2)= 
\Sigma_{h^0}^{\prime}(M_{h^0}^2) + 
(\delta Z_{\Phi_1}^{\overline{MS}}\sin^2\alpha +
\delta Z_{\Phi_2}^{\overline{MS}} \cos^2\alpha ) \, , 
\eeqa
in terms of the unrenormalized self-energy ${\Sigma}_{h^0}$.

The vertex counterterm $\delta(h^0 b\bar{b})$ in~(\ref{oneloopterms})
is given by
\beqa
\label{vertexCT}
\delta(h^0 b\bar{b}) = \frac{\delta m_b}{m_b} + \delta Z_V^b + 
\frac{\delta v_1}{v_1} \label{dhbb} \, ,
\eeqa
where $\delta m_b$ is the $b$-quark mass counterterm,
$\delta Z_V^b$ the $b$-quark field renormalization, and 
$\delta v_1/v_1$ the counterterm for the vacuum expectation value $v_1$. 
The $b$-quark is treated on-shell,
and $\delta Z_V^b$ is fixed by the condition that 
the residue of the $b$-quark propagator is normalized to unity;
consequently, we do not need external
wave-function renormalization for the $b$-quarks. 
In this way we obtain
\beqa
\frac{\delta m_b}{m_b} + \delta Z_V^b = \Sigma_S^b(m_b^2) - 2 m_b^2 
\left[ \Sigma_S^{\prime b}(m_b^2) + \Sigma_V^{\prime b}(m_b^2) \right] 
\eeqa
in terms of the scalar functions of the $b$-quark self-energy,
\beqa 
\Sigma^b(p) &=& \not{\!p}\,\Sigma_V^b (p^2) \,+\, 
                \not{\!p} \gamma_5\,\Sigma_A^b (p^2)\, + \,
                m_b\, \Sigma_S^b (p^2) \, .
\eeqa

In order to get the counterterm for $v_1$ in~(\ref{vertexCT}) we take 
over the condition of~\cite{dabelstein}, as formulated for the MSSM,
\vspace*{-0.1cm}  
\beqa
\label{vacren}
\frac{\delta v_2}{v_2} = \frac{\delta v_1}{v_1} \qquad ,\qquad \qquad
{\mbox{implying}} \qquad\qquad
\frac{\delta v_i}{v_i} = \frac{\delta v}{v}
\quad (i= 1,2) \, .
\eeqa
Since $v$ enters the gauge sector, $\delta v$
is related to charge and gauge-boson mass renormalization,
in the on-shell scheme given by
\beqa
2 \frac{\delta v}{v} & =  &
\cos^2\beta\, \delta Z_{\Phi_1}^{\overline{\rm MS}} + \sin^2\beta\,
\delta Z_{\Phi_2}^{\overline{\rm MS}} \nonumber \\ &+&
\Sigma_{\gamma\gamma}^{\prime}(0) +
2\frac{s_W}{c_W} \frac{\Sigma_{\gamma Z}(0)}{M_Z^2}
-\frac{c_W^2}{s_W^2} \frac{\Re\Sigma_{ZZ}(M_Z^2)}{M_Z^2} +
\frac{c_W^2-s_W^2}{s_W^2} \frac{\Re\Sigma_{WW}(M_W^2)}{M_W^2} \, .
\eeqa
It is interesting to note that, like in the MSSM, the difference 
$\frac{\delta v_2}{v_2}-\frac{\delta v_1}{v_1}$
is a UV-finite quantity. Moreover, the singular part of
$\delta v$ is identical to that of the MSSM case,
\beqa
\Bigm(\frac{\delta v}{v}\Bigm)_{\overline{\rm MS}} = 
-\frac{1}{4(4\pi)^2} (3 g^2 + g^{\prime 2}) \Delta \, .
\eeqa 

For a better physical understanding of the formal 
condition~(\ref{vacren}) it is enlightening to  consider
the ratio of the $A^0 \to d\bar{d}$ and $A^0 \to u\bar{u}$ decay widths,
which reads at the one-loop level (see also~\cite{fermionic})
\beqa
\frac{\Gamma_1(A^0\to d\bar{d})}{\Gamma_1(A^0 \to u\bar{u})} 
& \sim & \tan^4\beta \{ 1 - 4(\frac{\delta v_2}{v_2} -
\frac{\delta v_1}{v_1})\nonumber \\ 
& & + 2 ( \frac{\delta m_d}{m_d} + \delta Z_V^d - \frac{\delta m_u}{m_u} -
\delta Z_V^u) + V_1^{A^0d\bar{d}} - V_1^{A^0u\bar{u}} \}\label{1loop}\, ,
\eeqa
where $V_1^{A^0q\bar{q}}$ is the one-loop vertex correction to $A^0q\bar{q}$
vertex. 
The non-universal quantities $\frac{\delta m_q}{m_q} + \delta Z_V^q$
are sufficient to cancel the UV divergences from $V_1^{A^0q\bar{q}}$.
Consequently $\frac{\delta v_2}{v_2} -
\frac{\delta v_1}{v_1}$  is UV finite, and imposing 
condition~(\ref{vacren}) defines 
$\tan\beta$ at one loop through eq.~(\ref{1loop}).

For completeness, we list the couplings needed for this study.
In the limiting situation $\alpha\rightarrow \beta-\pi/2$, 
all the scalar couplings entering
the one-loop amplitude [Fig.~1, $d_2$] either vanish or 
reduce to their SM values except for $h^0H^+H^-$, $h^0H^0H^0$ 
and $h^0A^0A^0$, which are given, respectively, in eq.~(\ref{h0hphm}) and by
\beqa
g[h^0H^0H^0] & \approx & 
-\frac{g}{2M_W}\{ M_{h^0}^2 +2(M_{H^0}^2 -\lambda_5 v^2)  \}  \, ,
\nonumber \\ 
g[h^0A^0A^0] & \approx & 
-\frac{g}{2M_W}\{ M_{h^0}^2 +2(M_{A^0}^2 -\lambda_5 v^2)  \} \, .
\label{ha0a0}
\eeqa

In the THDM type II under consideration,
the neutral Higgs couplings to a pair of fermions 
normalized to $gm_f/(2M_W)$ are given by
\beqa
& & g[h^0 b\bar{b}]= -\frac{\sin\alpha}{\cos\beta}\cong 1  \quad , \quad
g[h^0 t\bar{t}]= \frac{\cos\alpha}{\sin\beta} \cong 1 \label{htt} \quad ,\\
& & g[H^0 b\bar{b}]= \frac{\cos\alpha}{\cos\beta}\cong \tan\beta \quad , \quad
g[H^0t\bar{t}]= \frac{\sin\alpha}{\sin\beta}\cong -{\rm cot}\beta
\label{Htt} \quad ,\\
& & g[A^0 b\bar{b}]= \gamma_5 \tan\beta \qquad , \qquad 
g[A^0 t\bar{t}] = \gamma_5 {\rm cot}\beta \label{Aff}\quad .
\eeqa
The charged Higgs coupling to fermions reads
\beqa
g[H^- \bar{b}t]=\frac{g}{\sqrt{2}M_W}
\{ m_t\, {\rm cot}\beta \frac{1+\gamma_5}{2} + 
m_b\, \tan\beta \frac{1-\gamma_5}{2} \} \label{htb}\,.
\eeqa

\section{Numerical results}
In the numerical evaluation of this section, 
we have parameterized the Higgs sector with the following seven
parameters \cite{AM}: 
the mass of the light CP-even Higgs boson, $M_{h^0}$;
the masses of the CP-odd, of the heavy neutral CP-even and
of the charged Higgs bosons, which are assumed to be degenerate, 
$M_{H^0}=M_{A^0}=M_{H^\pm}$; the mixing angles $\alpha$ and $\beta$ 
chosen to fulfill $\alpha=\beta-\pi/2$; 
the parameter of the discrete-symmetry breaking
of the Higgs potential, $\lambda_5$. When varying these parameters
we take into account the tree-level unitarity constraints
as derived in~\cite{abdesunit}. Such unitarity requirements 
put constraints on $M_{H^\pm}$, $M_{h^0}$, $\lambda_5$, and  $\tan\beta$.
Constraints on the charged Higgs-boson mass 
and $\tan\beta$ are also obtained from
experimental data on the decays $b\to s \gamma$
and $Z\to b\bar{b}$. It has been shown in~\cite{bsg} that 
for models of the type THDM-II, data on $b\to s \gamma$ give preference to
rather heavy charged Higgs particles, 
$M_{H^\pm} \gsim 200$ GeV for $\tan\beta\geq 1$ 
and even stronger for lower values of $\tan\beta$. 
From $Z\to b\bar{b}$ decays, strong constraints on  $M_{H^\pm}$ are
obtained in particular for small $0.5 <\tan\beta < 1$, yielding
$M_{H^\pm} > 200$ GeV at $99\%$ confidence level~\cite{zbb}.
In our study, since we are interested 
in the decoupling regime, we will assume
that the charged Higgs-boson mass is above 250 GeV.

\begin{figure}[t]
\begin{center}\vspace*{-2.0cm}\hspace*{-1.0cm}
\begin{tabular}{ccc}
\resizebox{5.5cm}{!}{\includegraphics{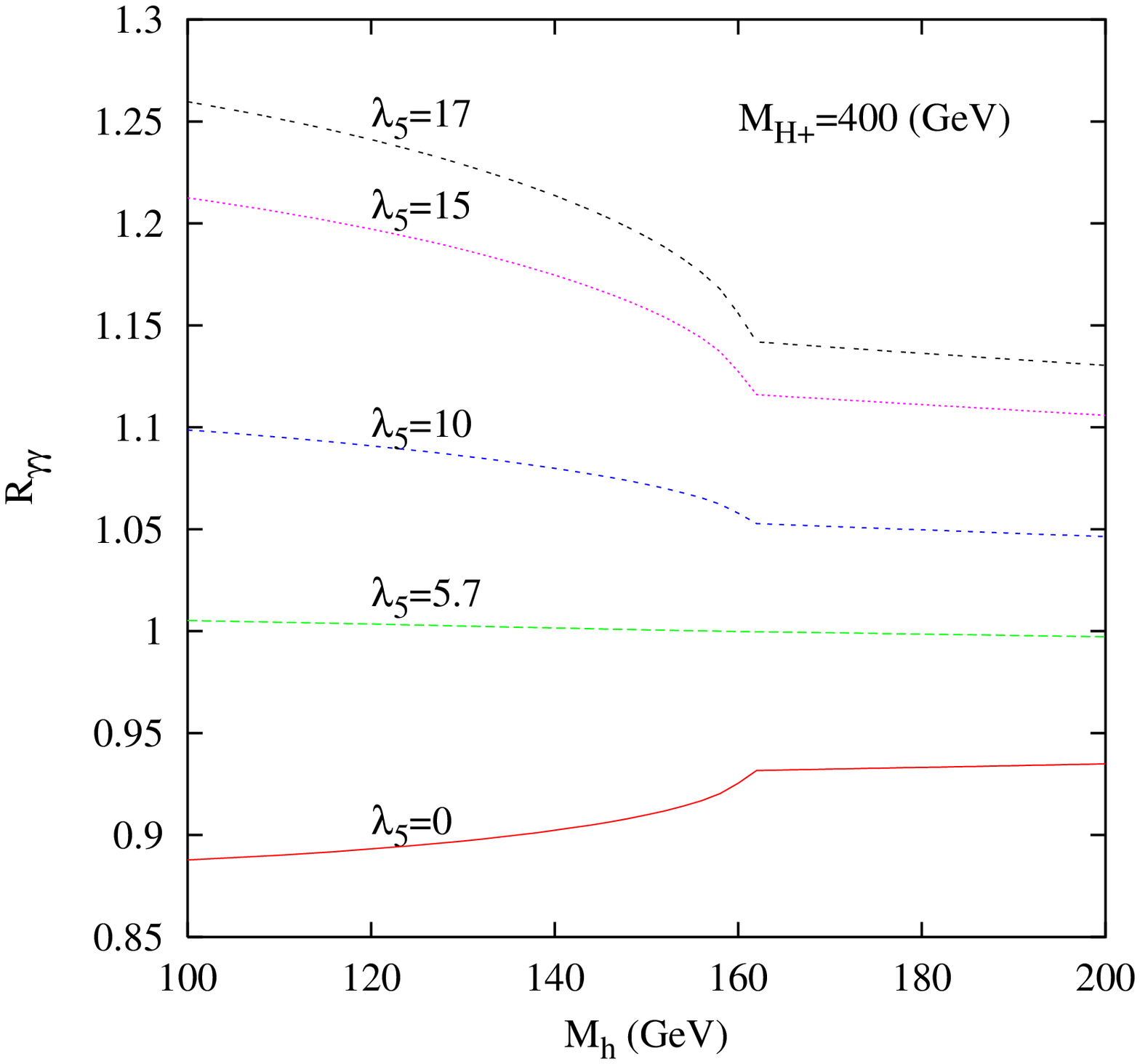}}&
\resizebox{5.5cm}{!}{\includegraphics{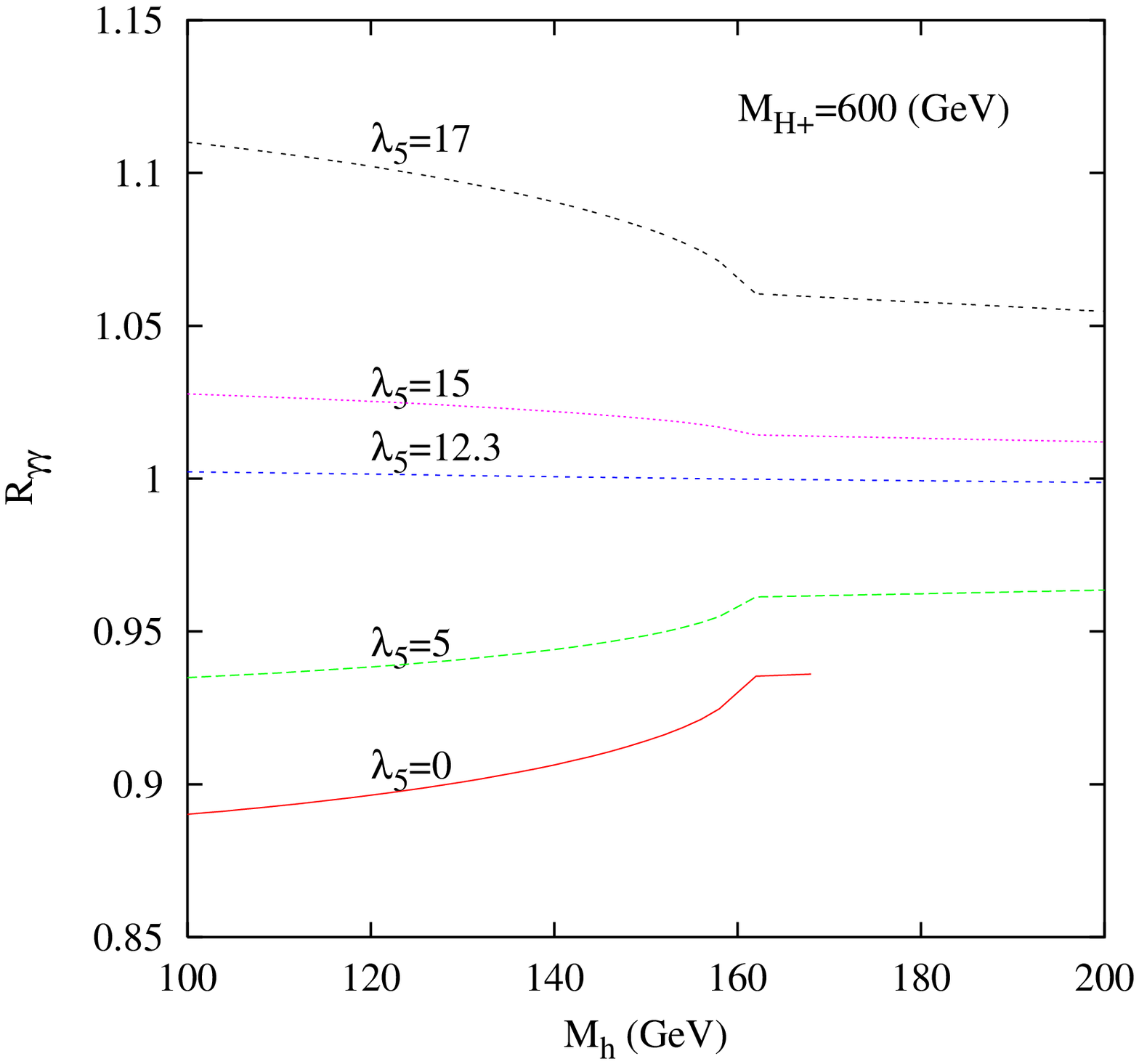}}&
\resizebox{5.5cm}{!}{\includegraphics{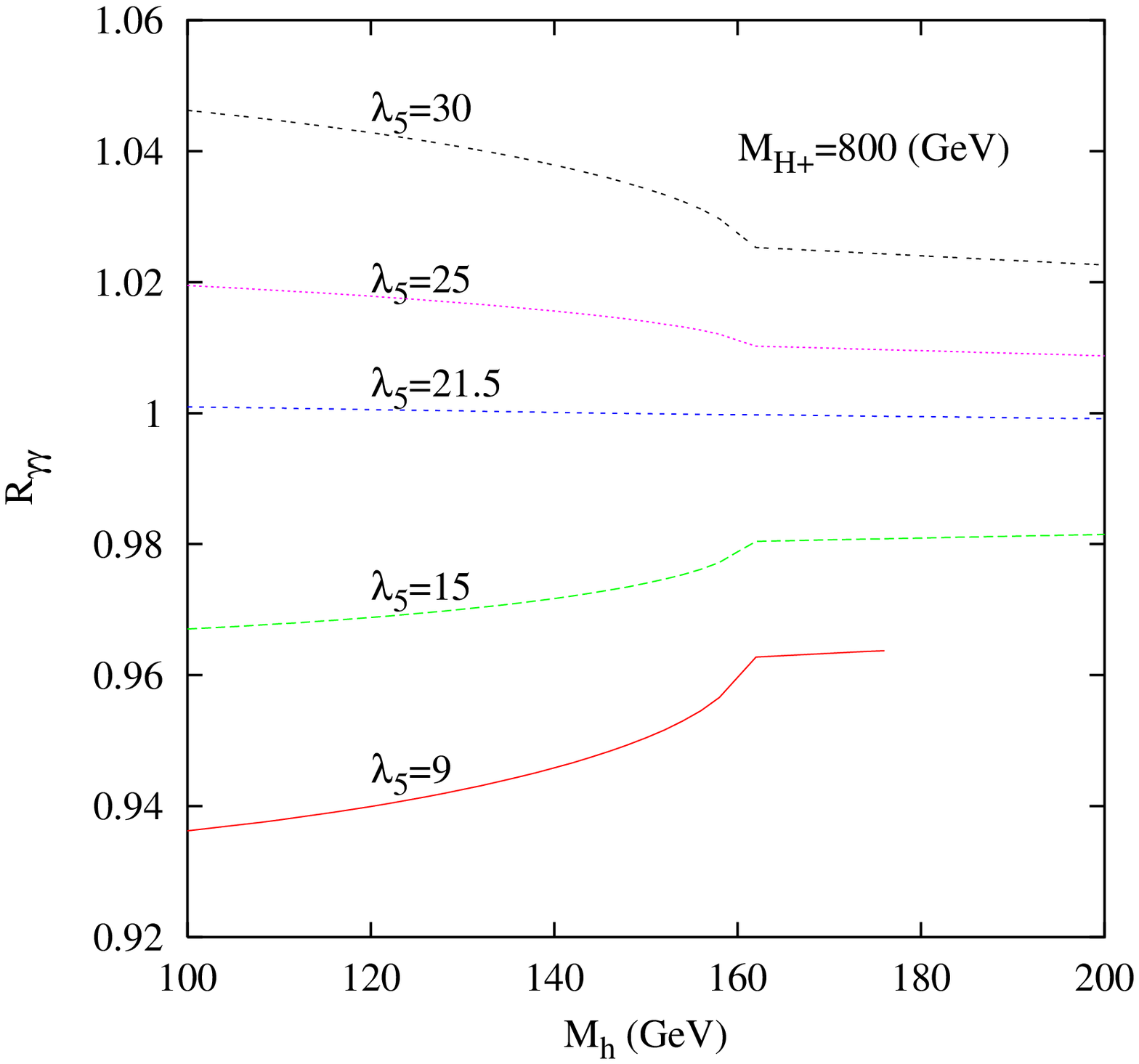}}\vspace*{-2.2cm}\\
\resizebox{5.5cm}{!}{\includegraphics{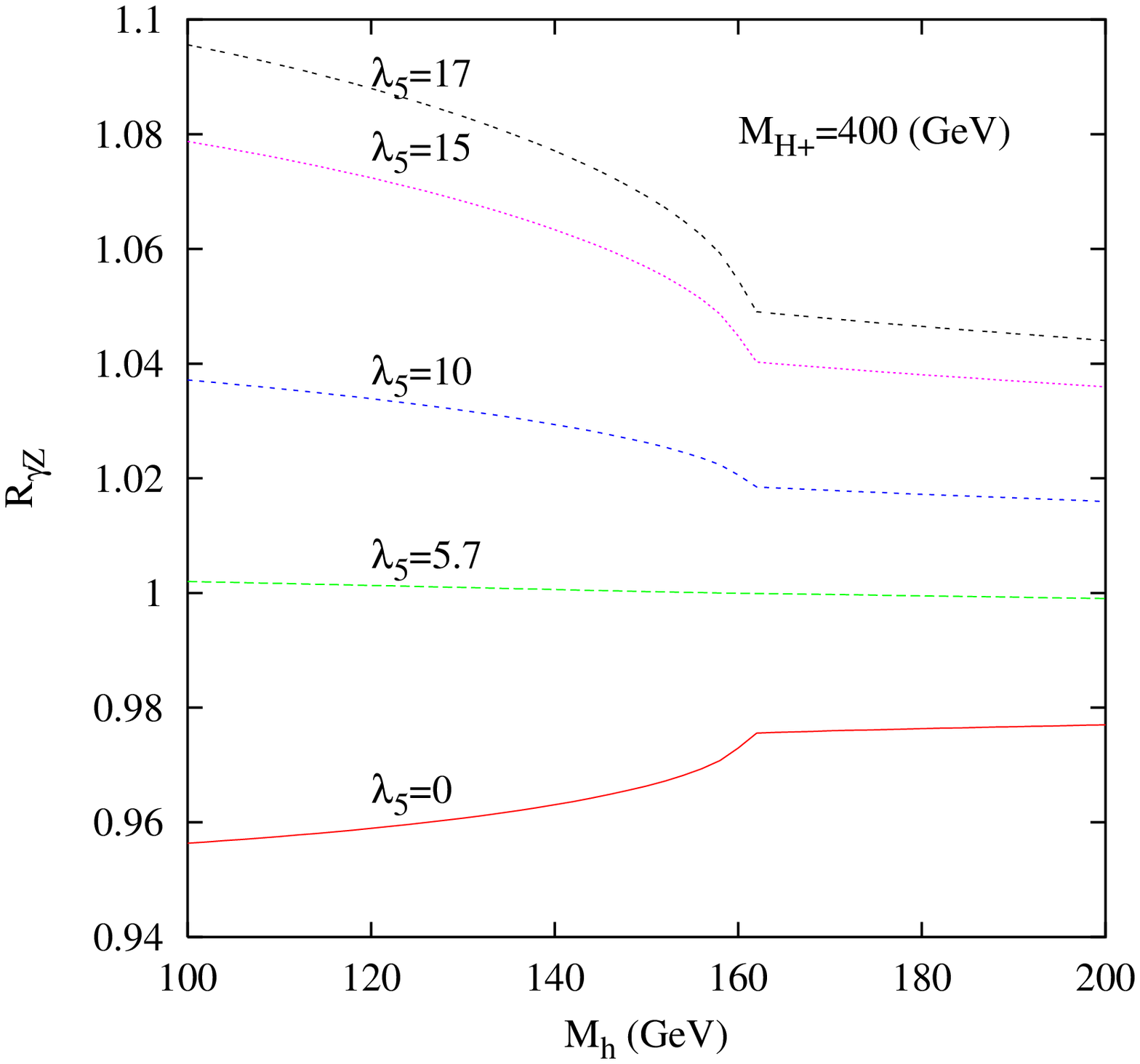}}&
\resizebox{5.5cm}{!}{\includegraphics{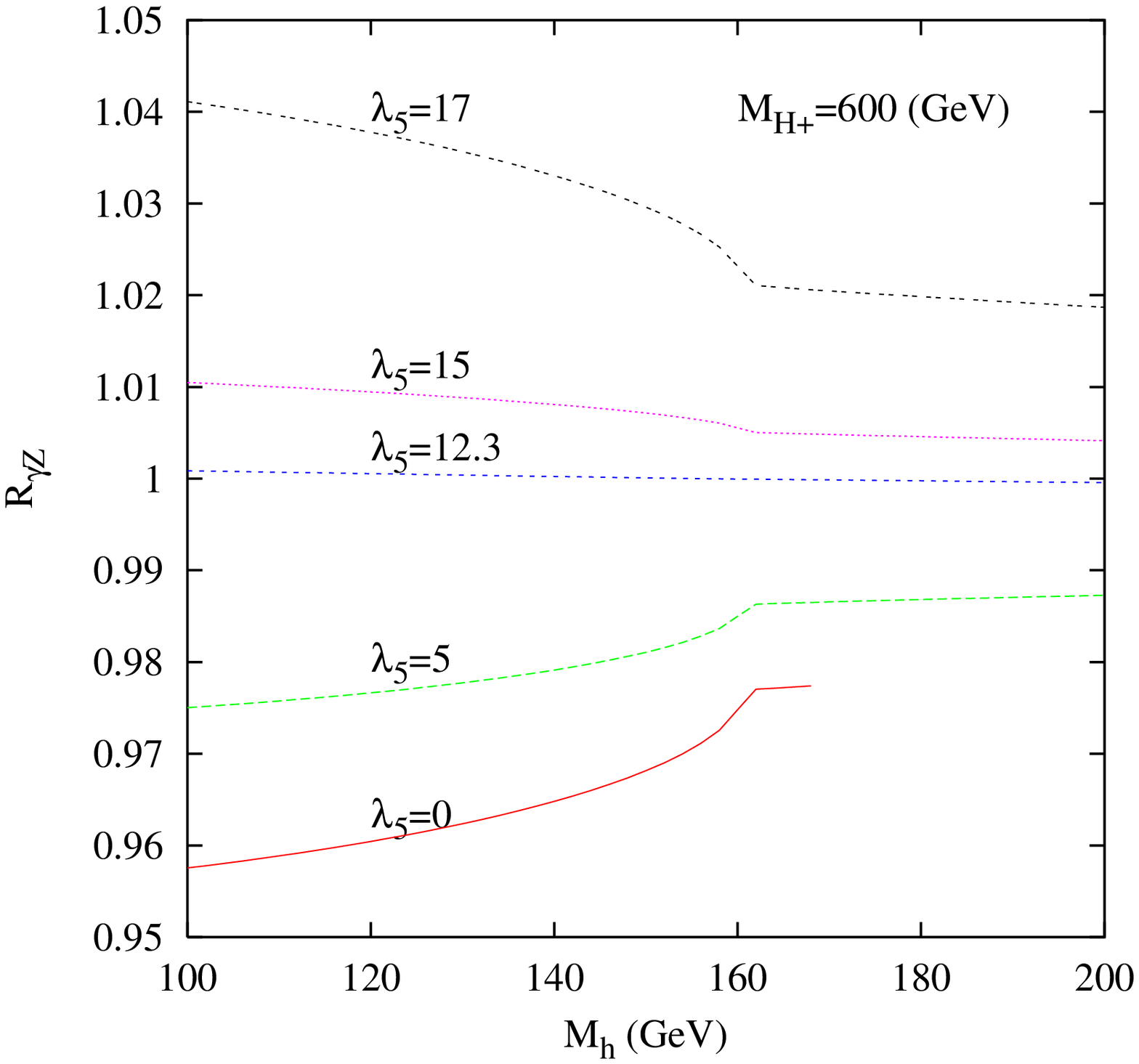}}&
\resizebox{5.5cm}{!}{\includegraphics{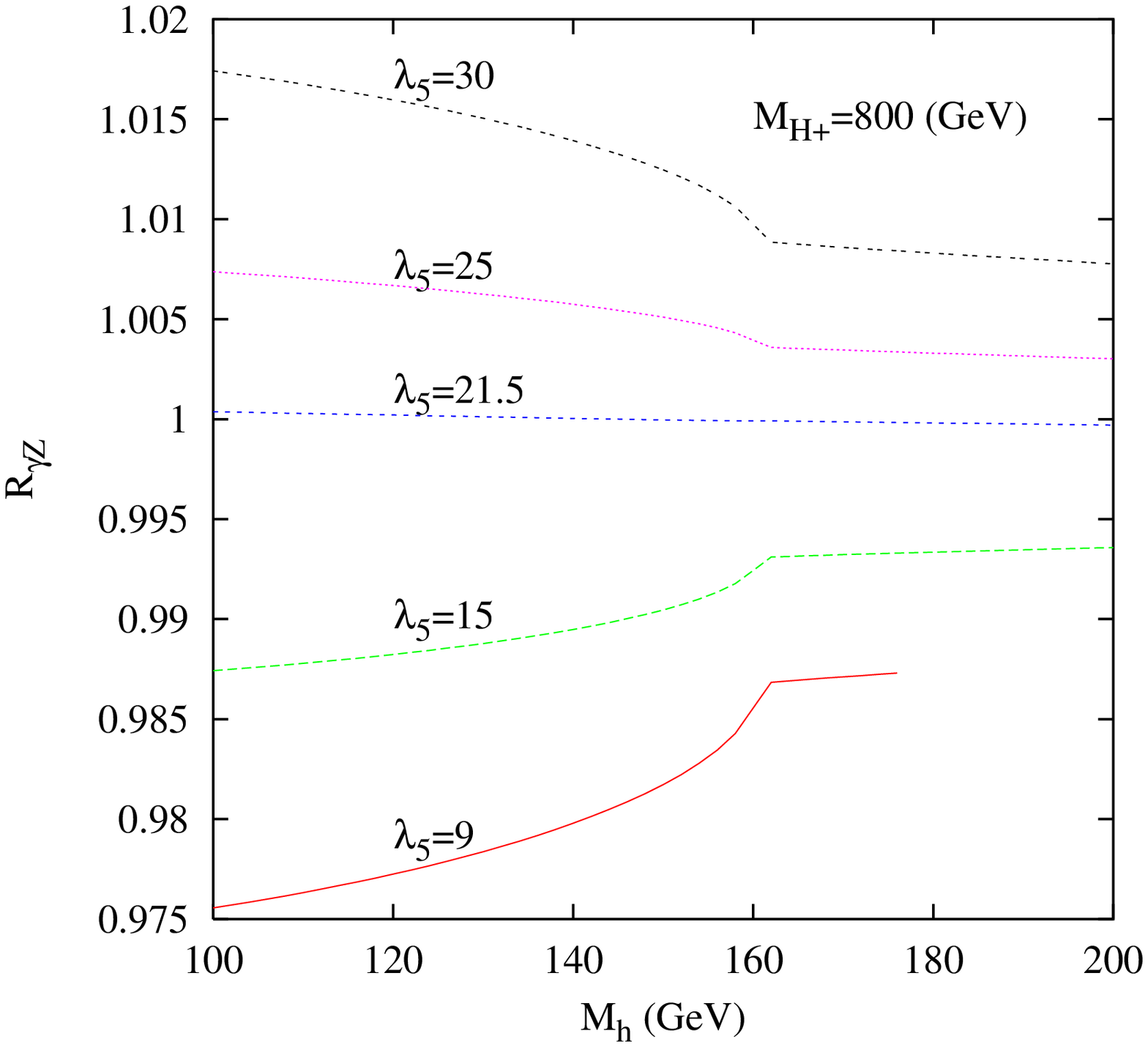}}\\
\end{tabular}
\end{center}\vspace*{-0.5cm}
\caption{$R_{\gamma\gamma}$ (upper plots) and $R_{\gamma Z}$ (lower
  plots) as function of $M_{h^0}$ for $M_{H^\pm}=400$ GeV (left plots), 
 $M_{H^\pm}=600$ GeV (middle plots), $M_{H^\pm}=800$ GeV (right plots)
 and $\tan\beta=1$.}
\label{fig:hV}
\end{figure}

For discussion of the non-standard effects, we introduce the following ratios,
\beqa
R_{\gamma V} =\frac{\Gamma(h^0\to \gamma V )}{\Gamma(H_{SM} \to \gamma V)} \, ,
\quad V=\gamma,Z, \ \  {\rm and} \ \ 
R_{bb} =\frac{\Gamma(h^0\to b\bar{b} )}{\Gamma(H_{SM} \to b\bar{b})} \, ,
\eeqa
which measure the deviations of the various $h^0$ partial widths from
their SM values with $M_{H_{SM}} = M_{h^0}$.
In Fig.~\ref{fig:hV}, these ratios are displayed for three values of 
the charged Higgs mass, for several values of $\lambda_5$, and for 
$\tan\beta \approx 1$. The choice for $\tan\beta$ is 
motivated by unitarity constraints;
the parameter space allowed by unitarity is large
for $\tan\beta \approx {\cal O}(1)$ and is reduced 
for large $\tan\beta$.\\
In the decoupling regime, the tree level 
couplings $h^0q\bar{q}$ become nearly equal to their SM values, 
and so the ratios $R_{\gamma V}$ are independent of $\tan\beta$.
  
As one can see in  Fig.~\ref{fig:hV}, for each value of
$M_{H^\pm}=400,600,800$ GeV one can find corresponding values of 
$\lambda_5^0\approx 5.7,12.3, 21.5$
\footnote{These values of $\lambda_5$ 
are obtained from eq.~(\ref{zero}) with $M_{h^0}=150$ GeV},
for which the charged-Higgs contribution vanishes,
thus yielding $R_{\gamma V} \approx 1$.
For $\lambda_5$ greater or smaller than the critical values
$\lambda_5^0$, the deviation in the partial width for $h^0\to {\gamma \gamma}$ 
can be as large as $25\% \,( 11\%)$ for $M_{H^\pm}=400$ GeV (600 GeV). 
For $M_{H^\pm}=800$ GeV, unitarity is violated
for $0<\lambda_5<9$ and the deviation in $h^0 \to {\gamma \gamma}$ 
can be only of the order $\pm 6\%$. We note that our results
agree with the results of \cite{maria}, but some of the parameters chosen in 
\cite{maria} violate the tree-level unitarity constraints.
For the ratio $R_{\gamma Z}$, the effect is not so large.
The charged-Higgs contribution in this case is decreased by the gauge 
coupling, which for the $Z$ is smaller than for the photon, 
$$\frac{g_{ZH^+H^-}}{g_{\gamma H^+H^-}}=
\frac{g_{\gamma ZH^+H^-}}{g_{\gamma\gamma H^+H^-}}=
\frac{(1-2s_W^2)}{(2s_Wc_W)}\approx 0.64\, . $$
Only for $M_{H^\pm}=400$ GeV and large $\lambda_5=17$, the 
deviation of the ratio 
$R_{\gamma Z}$ from unity is about 10\%.

\begin{figure}[t]
\begin{center}\vspace*{-2.0cm}\hspace*{-1.0cm}
\begin{tabular}{ccc}
\resizebox{5.5cm}{!}{\includegraphics{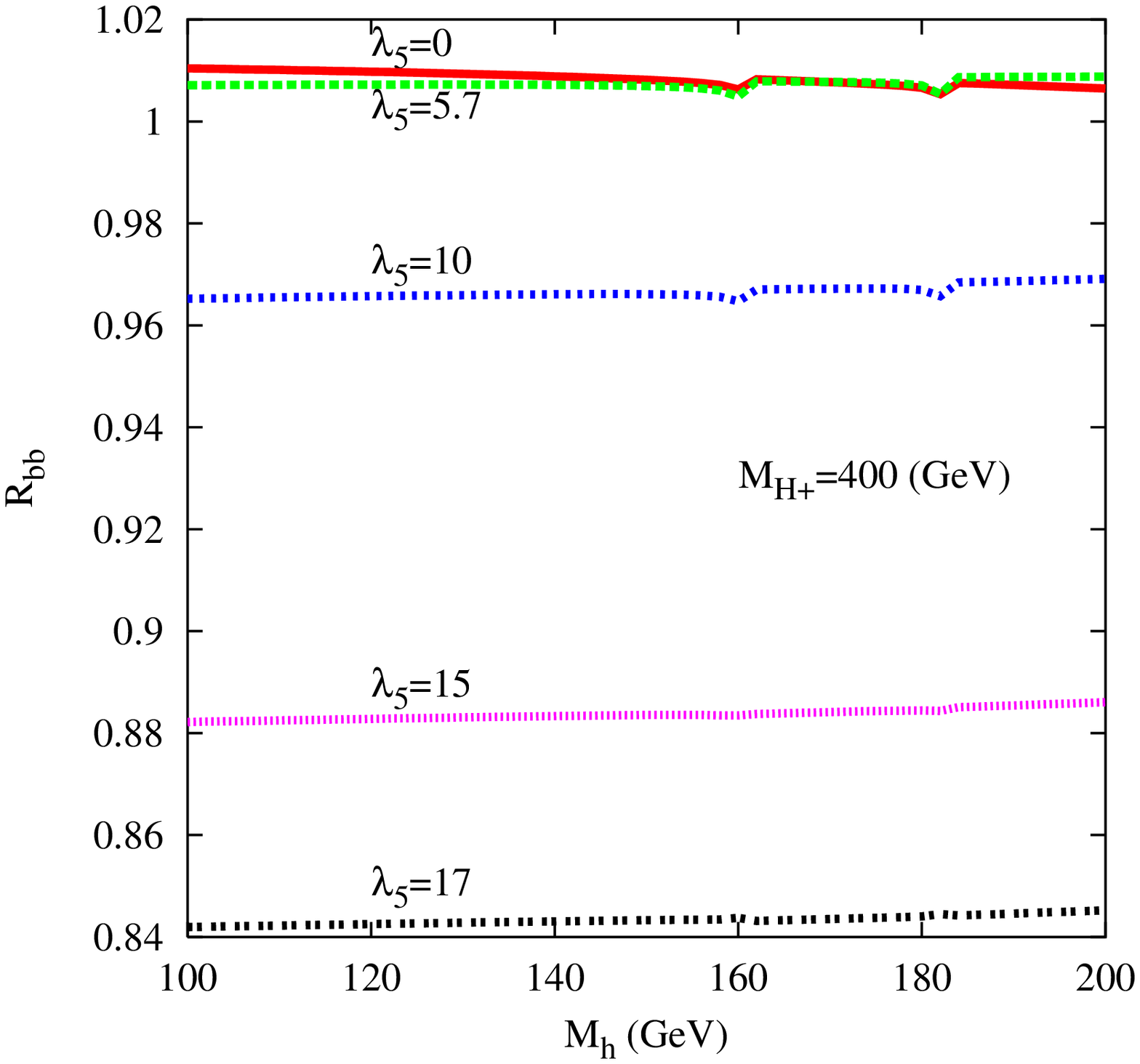}}&
\resizebox{5.5cm}{!}{\includegraphics{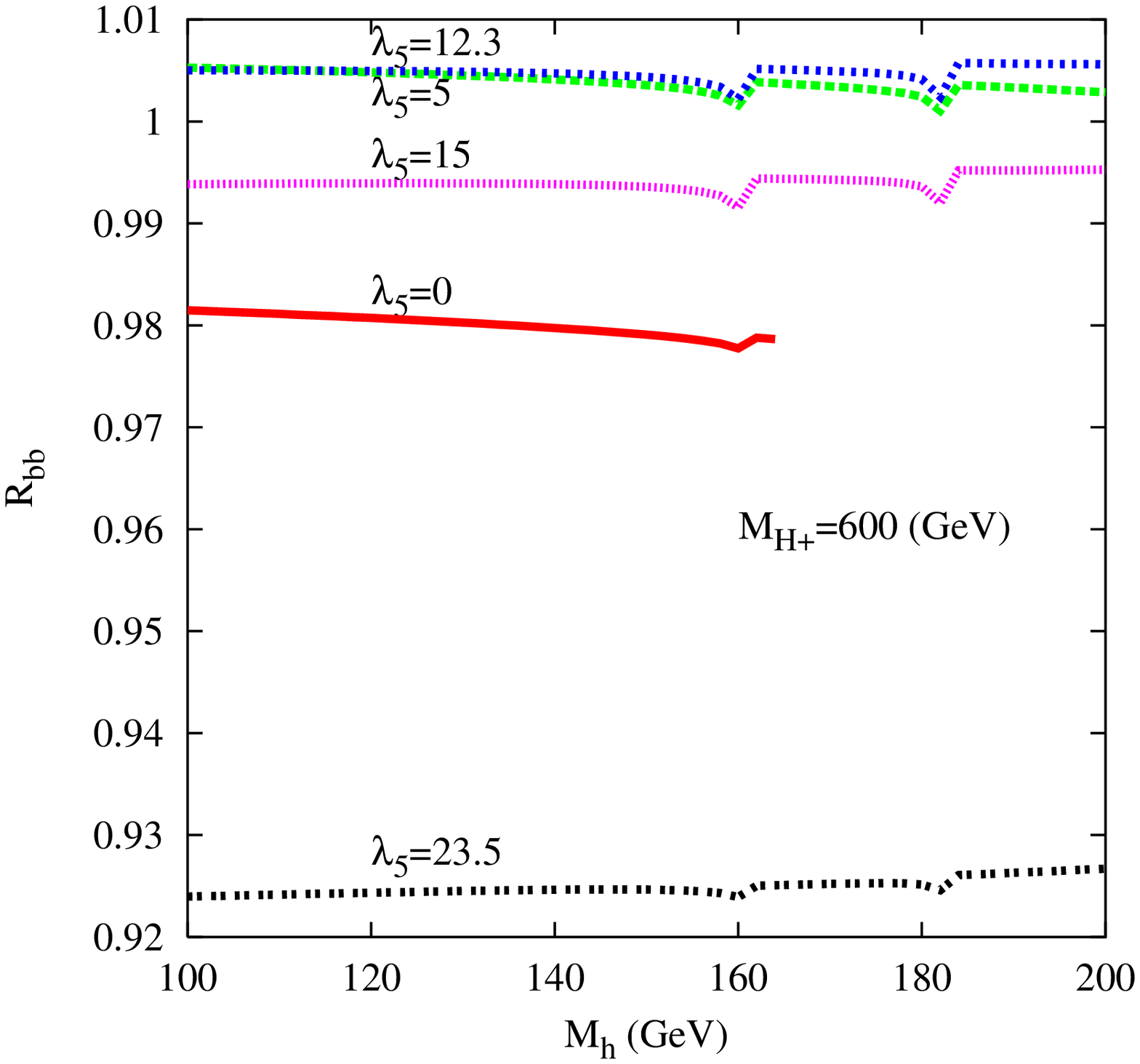}}&
\resizebox{5.5cm}{!}{\includegraphics{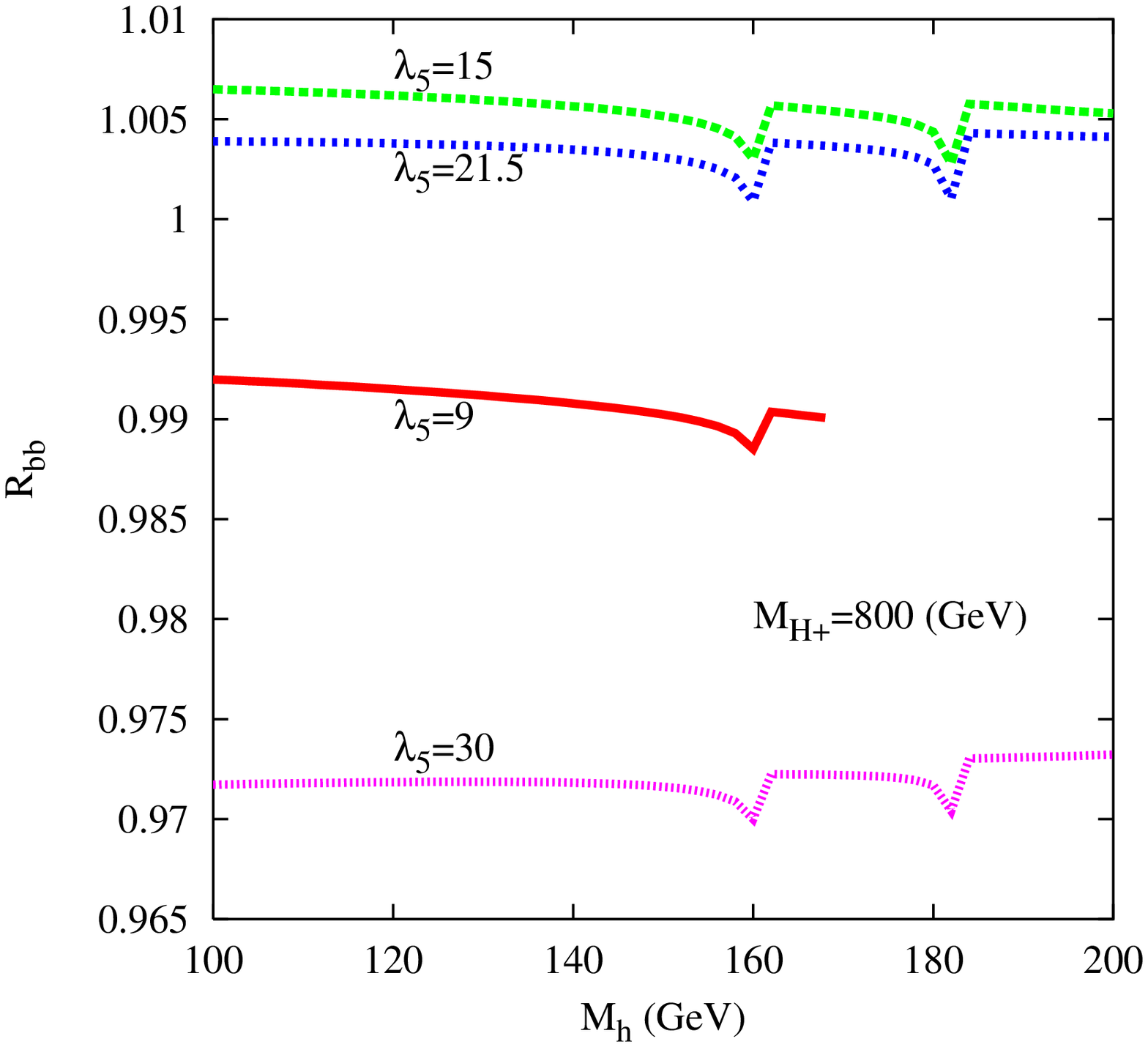}}\\
\end{tabular}
\end{center}\vspace*{-0.5cm}
\caption{$R_{bb}$ as function of $M_{h^0}$ 
for $M_{H^\pm}=400$ GeV (left), 
$M_{H^\pm}=600$ GeV (middle), $M_{H^\pm}=800$ GeV (right) 
and $\tan\beta=1$.}
\label{fig:hbb}
\end{figure}

For the Higgs-boson decay into a $b$-quark pair,
$h^{0} \to b \bar{b}$,  before discussing our 
numerical results, we would like to mention that we have 
done the following checks: \\
i) for the SM case $H_{SM} \to b \bar{b}$, we have reproduced the SM results
in perfect agreement with~\cite{HDK}, 
ii) for the THDM case, we have checked numerically that
we recover the SM corrections to $H_{SM}\to b \bar{b}$ 
in the decoupling scenario with all pure THDM couplings set to zero.

\begin{figure}[t]
\begin{center}\vspace*{-2.4cm}\hspace*{-1.0cm}
\begin{tabular}{ccc}
\resizebox{5.5cm}{!}{\includegraphics{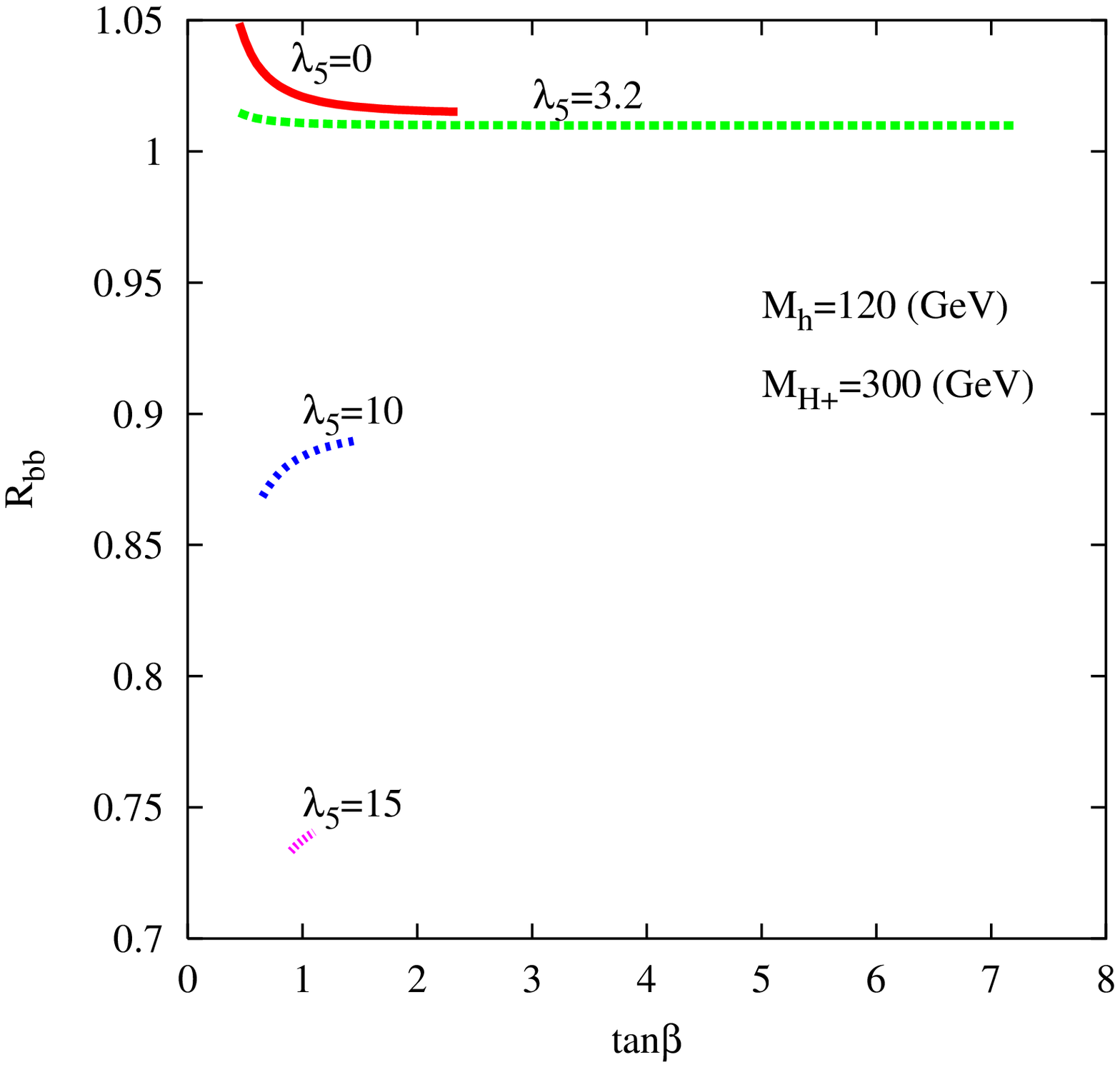}}&
\resizebox{5.5cm}{!}{\includegraphics{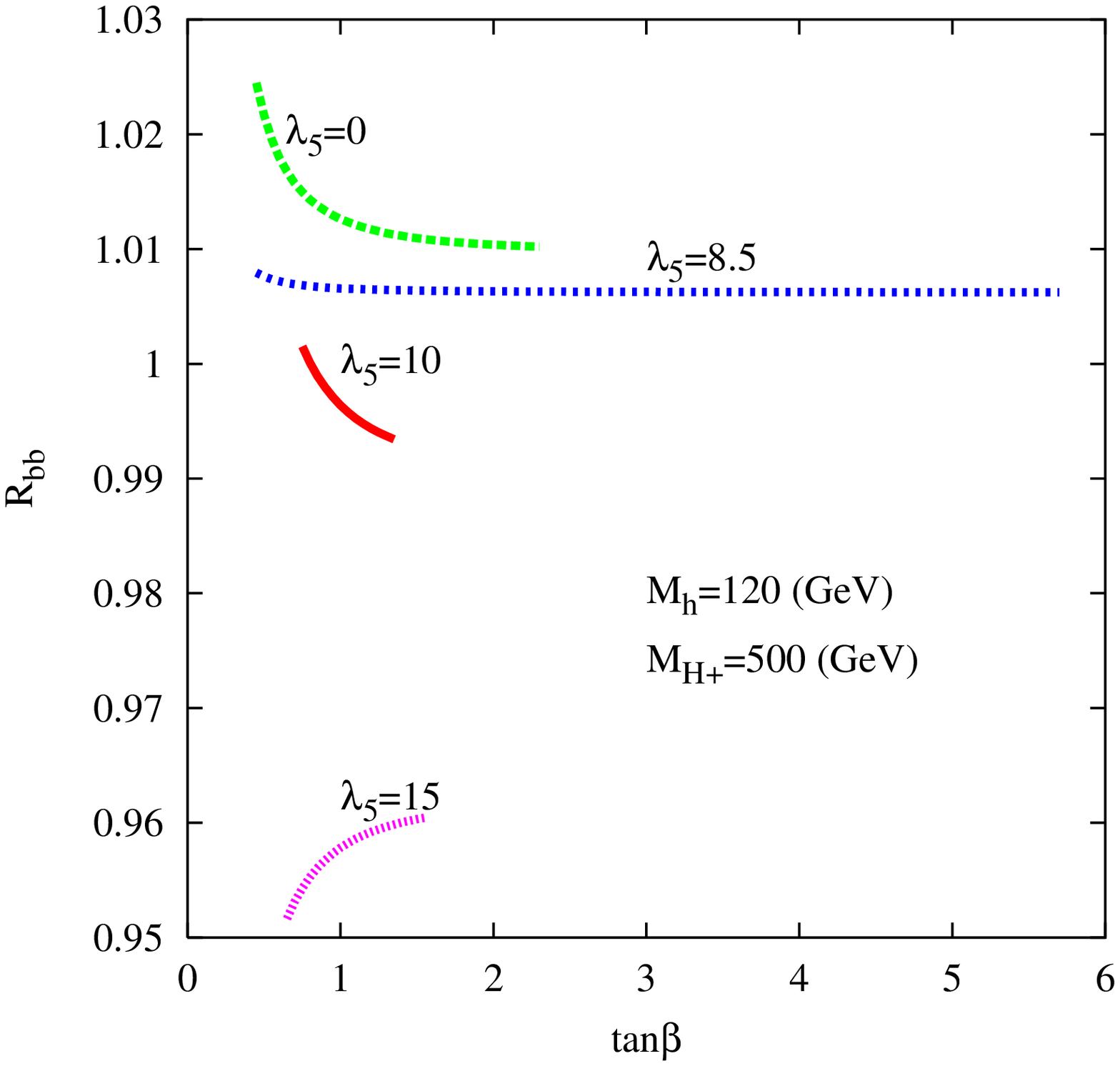}}&
\resizebox{5.5cm}{!}{\includegraphics{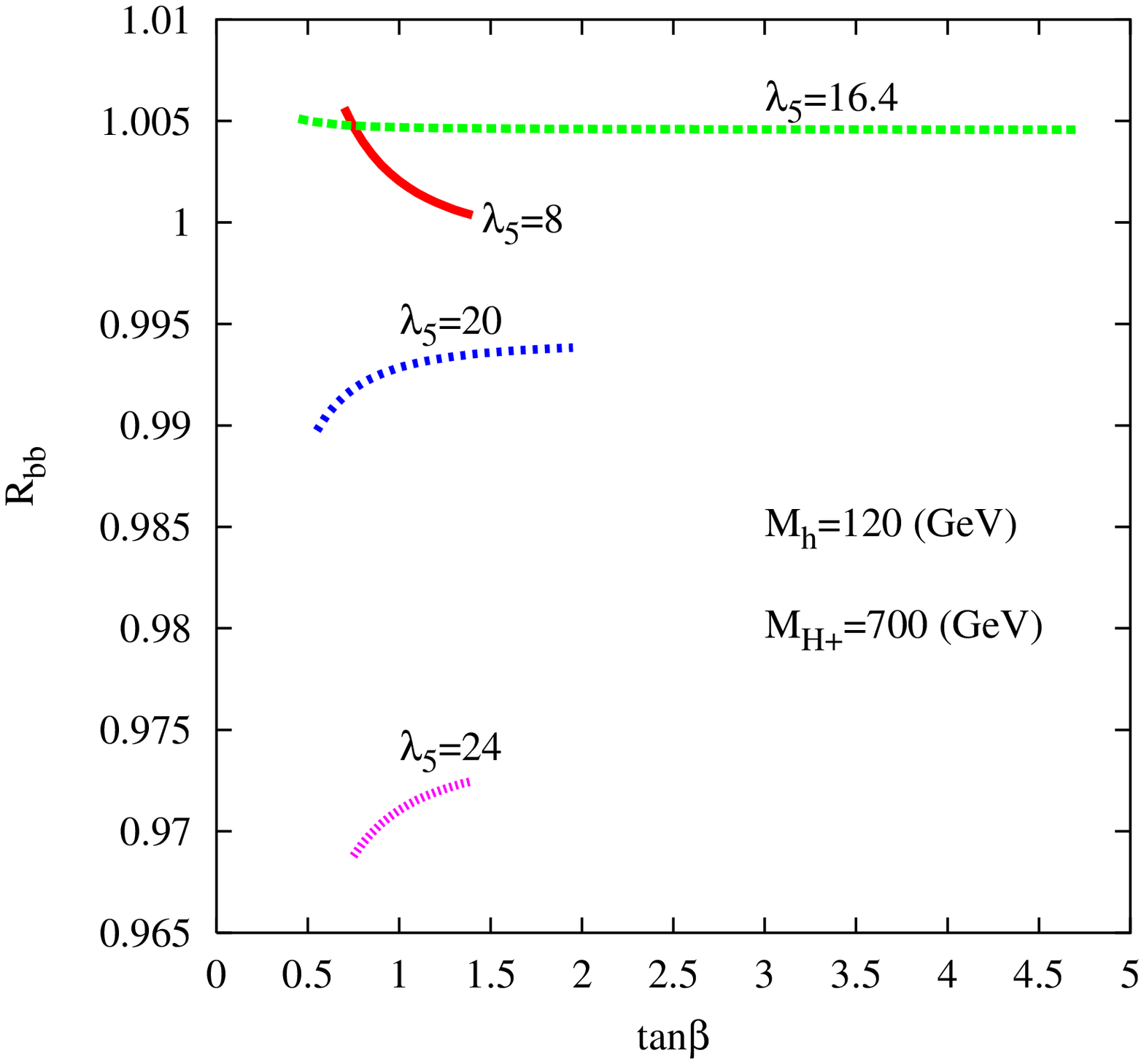}}\\
\end{tabular}
\end{center}\vspace*{-0.5cm}
\caption{$R_{bb}$ as function of $\tan\beta$ for 
$M_{H^\pm}=300$ GeV (left), 
$M_{H^\pm}=500$ GeV (middle), $M_{H^\pm}=700$ GeV (right)
 and $M_{h^0}=120$ GeV.}\vspace*{-0.1cm}
\label{fig:hbb1}
\end{figure}

As explained in the previous section, the Higgs fields are renormalized in 
the $\overline{\rm MS}$ scheme and hence the result depends on  
the renormalization scale $\mu_{\overline{\rm MS}}$. This dependence, however, 
is rather weak for $M_Z \leq \mu_{\overline{\rm MS}} \leq M_t$, as has
been checked explicitly.
In the following we will use $\mu_{\overline{\rm MS}}=M_Z$.

In Fig.~\ref{fig:hbb}, we illustrate the ratio
$R_{bb}$ as a function of $M_{h^0}$ for the same parameters 
as applied in the $R_{\gamma V}$ cases. It is clear from this plot 
that sizeable effects also appear 
in $h^0\to b\bar{b}$, on equal footing as in 
$h^0\to \gamma\gamma$ and $h^0\to \gamma Z$.
They originate from two sources, allocated to 
diagrams $d_1$ and $d_2$ of Fig.~\ref{fig:diagrams}.
The effects from $d_1$ are less than 
1\%, while large effects  are due to $d_2$. 
It is interesting to observe that for large values of $\lambda_5$ the THDM
contribution to $h^0\to b\bar{b}$ is reduced with respect to the SM
($R_{bb}<1$). For $M_{H^\pm}=400$ GeV ($600$ GeV), the deviation of 
$\Gamma(h^0\to b\bar{b})$ from the SM value can reach 16\% for
$\lambda_5=17$ (resp 8\% for $\lambda_5=23.5$).
For heavy charged Higgs bosons, like $M_{H^\pm}=800$ GeV,
the deviation is about 3\% for $\lambda_5=30$.

In the decoupling limit, the $b$-quark mass gets a factor
$\tan\beta$  in either the  $H^0b\bar{b}$, $A^0b\bar{b}$, and $H^-\bar{b}t$
couplings
(see eqs.~(\ref{Htt}), (\ref{Aff}) and (\ref{htb})), while the top mass
gets a factor ${\rm cot} \beta$. Consequently the top (bottom) 
effect is enhanced for small (large) $\tan\beta$. 
We have studied the sensitivity to $\tan\beta$, but 
the unitarity requirements impose severe constraints on $\tan\beta$, 
which effectively turns out to be of the order one. 
Note that for $\lambda_5 \simeq \lambda_5^0$ no such constraint exists.
In Fig.~\ref{fig:hbb1}, the ratio $R_{bb}$ is displayed as a 
function of $\tan\beta$ for $M_{h^0}=120$ GeV, 
three values of  $M_{H^\pm}=300$, 500,  700 GeV,
and several values of $\lambda_5$. For every value of $M_{H^\pm}$
we have chosen $\lambda_5$ in such a way that the couplings 
(\ref{h0hphm}) and (\ref{ha0a0}) vanish, according to (\ref{zero}), 
and consequently the diagram $d_2$ is zero.
For $M_{h^0}=120$ GeV and $M_{H^\pm}=300$, 500, 700 GeV,
the zero of (\ref{zero}) is located at $\lambda_5=3.2$, 8.5, 16.4, 
respectively. For those values, the deviation of $h^0\to b\bar{b}$
from the SM width arises mainly from diagram $d_1$; 
$R_{bb}$  is not very sensitive to $\tan\beta$, with  at most
1\%  deviations for $0.5 \leq \tan\beta \leq 7$.
For the other values of $\lambda_5$ the effect can be large,
but $\tan\beta$ is significantly restricted around unity.

\begin{figure}[t]
\begin{center}
\begin{tabular}{cc}
\resizebox{8.1cm}{!}{\includegraphics{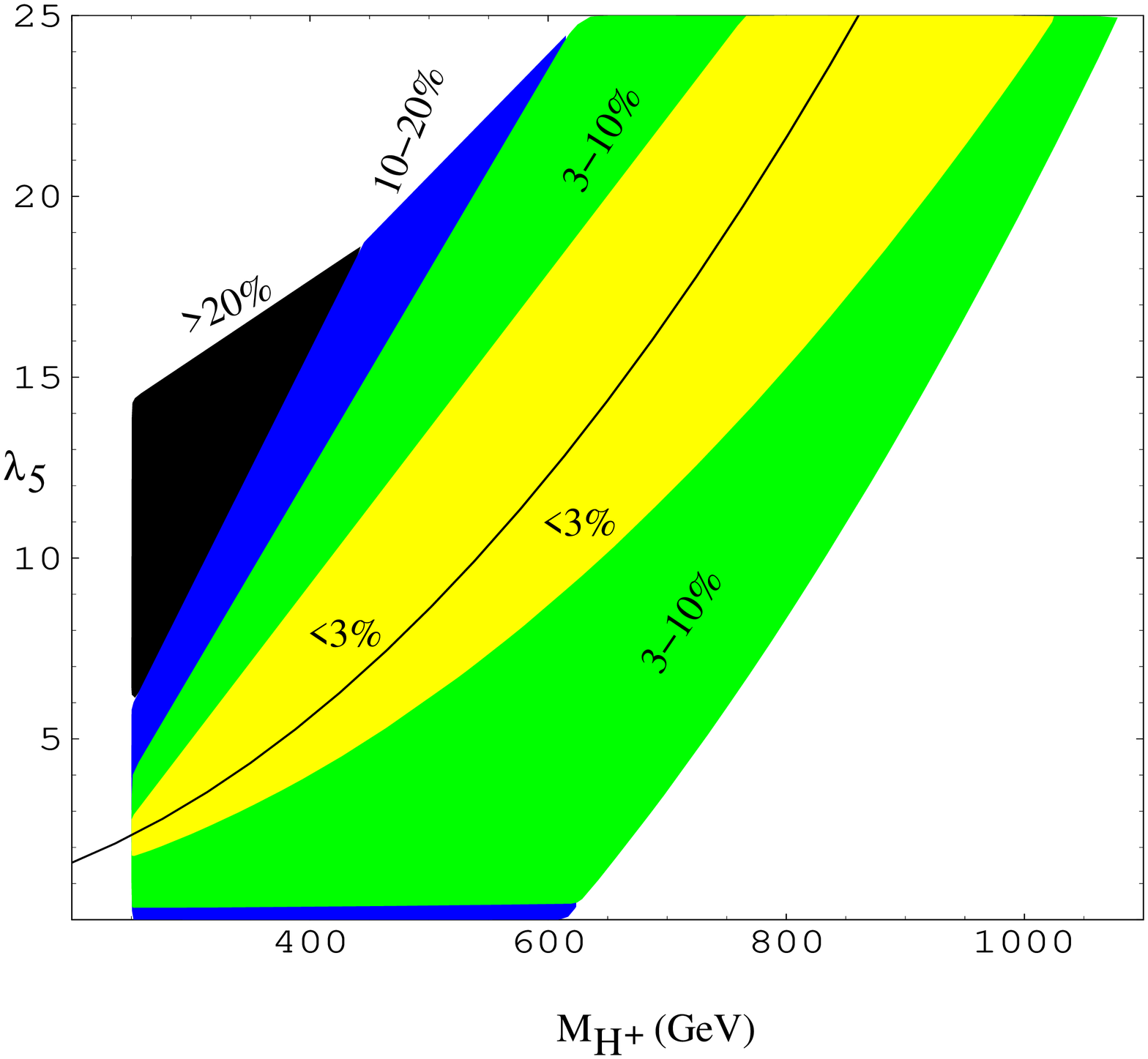}}&
\resizebox{8.0cm}{!}{\includegraphics{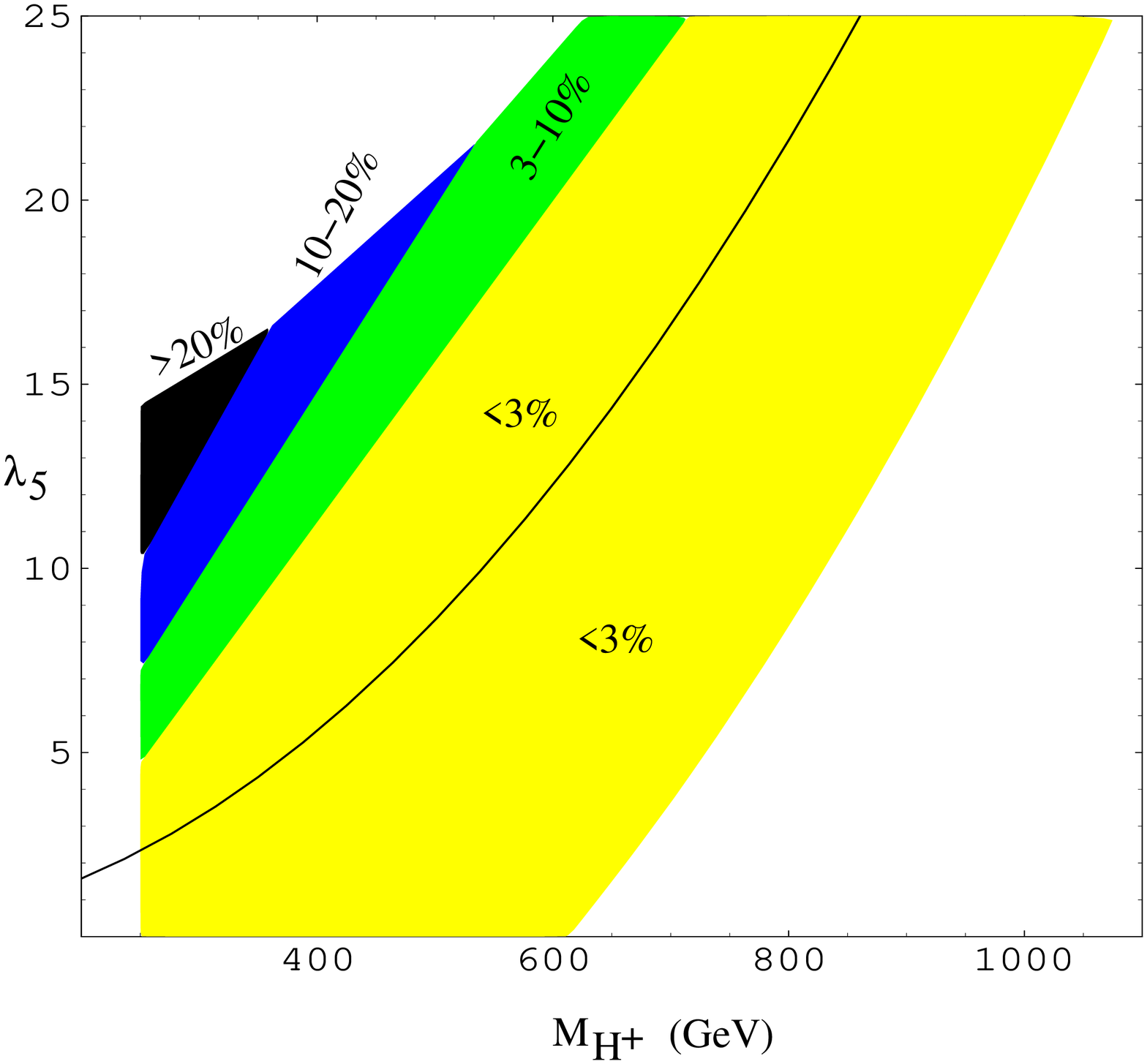}}\\
\end{tabular}
\end{center}\vspace*{-0.5cm}
\caption{Contours for $\Delta_{\gamma\gamma}=|R_{\gamma\gamma}-1|$
(left panel), $\Delta_{bb}=|R_{bb}-1|$
(right panel) in the $(M_{H^\pm},\lambda_5)$ plane for $M_{h^0}=120$ 
GeV, $\tan\beta=1$ and $\mu_{\overline{\rm MS}}=M_Z$} 
\label{fig:fig2}
\end{figure}

In Fig.~\ref{fig:fig2}, we show a contour plot for 
$\Delta_{\gamma\gamma}=|R_{\gamma\gamma}-1|$  (left panel)
and  $\Delta_{bb}=|R_{bb}-1|$ (right panel)
in the plane  $(M_{H^\pm},\lambda_5)$ for $\tan\beta=1$ and $M_{h^0}=120$ GeV.
The light-grey (yellow) area shows the region where the deviation
from the SM value is less than $\pm 3\%$. 
The solid black line in that region indicates the situations 
for which the couplings (\ref{h0hphm}) and (\ref{ha0a0}) vanish. In these cases
the deviation of the THDM decay widths for $h^0\to b\bar{b}$ from 
the corresponding SM value is less than 1\%.

Concerning the left panel, $\Delta_{\gamma\gamma}$, 
away from the light-grey (yellow) region and for charged Higgs-boson masses 
$250 \lsim M_{H^\pm} \lsim 600$ GeV and $\lambda_5\gsim 3$,
the deviations are greater than 3\% and can become as large as 
20\% for $M_{H^\pm}\lsim 400$ GeV and $\lambda_5\gsim 6$.
One can also see that for small $\lambda_5<1$ and $M_{H^\pm}$
below  600 GeV, the deviation is in the range $10 - 20 \%$.
For $\Delta_{bb}$, the behaviour is similar.
For $250 \lsim M_{H^\pm} \lsim 700$ GeV and $\lambda_5\gsim 5$, the
deviations from the corresponding SM value exceed $3\%$;
for $M_{H^\pm}\lsim 400$ GeV and $\lambda_5\gsim 10$, effects
above $20\%$ are found.

\section{Conclusions}

We have studied the Higgs-boson decays $h^0\to \gamma \gamma$, 
$\gamma Z$ and $b\bar{b}$ in the THDM within the 
decoupling-regime scenario. We have shown that, 
even when taking into account unitarity constraints in the scalar sector,
both in $h^0\to \gamma \gamma$ and $h^0\to \gamma Z$
large quantum effects can occur, i.e we have found sizeable
differences between these decays and the corresponding 
decay widths of the SM Higgs boson. For charged Higgs 
bosons not too heavy, of about $400$ GeV, the deviations in 
the decay width for $h^0\to \gamma \gamma$ ($h^0\to \gamma Z$) from
the corresponding SM values can reach the order of 25\% (10\%).

For the dominant Higgs-boson decay into $b$-quark pairs,
the partial width $\Gamma(h^0\to b\bar{b})$ in the decoupling limit
is identical to the SM one at tree level.
It turns out that for the same scenarios leading to sizeable
effects in the loop-induced bosonic decays, 
significant quantum effects are also present
in $\Gamma(h^0\to b\bar{b})$ at one-loop order.
Those effects originate  mainly from the scalar self-couplings
$h^0H^\pm H^\mp$, $h^0H^0H^0$ and $h^0A^0A^0$. For certain regions 
of the charged Higgs-boson mass and the parameter 
$\lambda_5$, the deviation with respect to the SM value can be also 
of the order of 25\%.
Hence, quantum effects in $h^0\to b\bar{b}$ can be of the same
size as in the $\gamma\gamma,\gamma Z$ decay modes. Therefore, not only the 
one-loop effects to $h^0\to \{\gamma\gamma ,\gamma Z \}$
but also the quantum contributions  to $h^0\to b\bar{b}$ 
can be used to distinguish between THDM and SM.

At the end, we would like to emphasize, that for large values of the 
charged Higgs boson masses, unitarity requires large $\lambda_5$ 
(or large $\mu_{12}^2$). In this decoupling limit \cite{hojon}, 
and as it can be seen from the figures, 
the deviations of the observables we have been considering above 
from their SM values are very small.
                                           
\section*{{\large{Acknowledgment:}}}
This work was supported in part by the European Community's Human
Potential Programme under contract HPRN-CT-2000-00149 
``Physics at Colliders''. 
A. Arhrib acknowledges the Alexander von Humboldt Foundation.
MCP thanks the Werner-Heisenberg-Institut for the kind
hospitality during his visit where part of this work was done.
We are grateful to T. Hahn and O. Brein for computing assistance
and useful discussions.

\end{document}